\documentclass[a4paper,12pt]{article}
\usepackage[cp1251]{inputenc}
\usepackage{epsfig}
\usepackage[english]{babel}
\usepackage{amssymb}

\begin{document}
\title{Siegel disk for complexified H\'{e}non map}
\author{O.B. Isaeva, S.P. Kuznetsov}
\date{}
\maketitle\begin{center} \emph{Institute of Radio-Engineering and
Electronics of
RAS, Saratov Branch, \\ Zelenaya 38, Saratov, 410019, Russia \\
E-mail: IsaevaO@rambler.ru}\end{center}

\begin{abstract}
It is shown that critical phenomena associated with Siegel disk, 
intrinsic to 1D complex analytical maps,
survives in 2D complex invertible dissipative H\'{e}non map.
Special numerical method of estimation of the Siegel disk 
scaling center position (for 1D maps it corresponds to extremum) 
for multi-dimensional invertible maps are developed.
\end{abstract}

It is known that complexification of real 1D logistic map
\begin{equation}\label{log}
z_{n+1}=f(z_n)=\lambda-z_{n}^{2},
\end{equation}
where $\lambda,z\in\mathbb{C}$ leads to the origination of the
Mandelbrot set at the complex parameter $\lambda$
plane~\cite{Peitgen} and a number of other accompanying phenomena.

Opportunity  for realization of the phenomena, characteristic for
the dynamics of complex maps (Mandelbrot and Julia sets etc.) at the physical
systems seems to be interesting problem~\cite{Beck,Ipnd,IP,I1,I2}. In the
context of this problem the following question is meaning: Does
phenomena of dynamics of the 1D complex maps (like classic
Mandelbrot map~(\ref{log})) survive for the more realistic 
(from the point of view of possible physical applications) model 
-- two-dimensional maps invertible in time.
For example, more realistic model rather than logistic map, is
the H\'{e}non map
\begin{equation}\label{hen}
z_{n+1}=f(z_{n},w_{n})=\lambda-z_{n}^{2}-d\cdot w_{n},\qquad
w_{n+1}=g(z_{n},w_{n})=z_{n}.
\end{equation}
In the real variable case the system~(\ref{hen}) is 2D invertible
map and, hence, can be realized as Poincare cross-section of flow
system with three dimensional phase space -- minimal dimension,
providing opportunity of nontrivial dynamics and chaos. H\'{e}non
map is suitable for modelling of the chaotic dynamics of the
generator with non-inertial nonlinearity, dissipative oscillator
and rotator with periodic impulse internal force
etc.~\cite{Kuznetsov}. Moreover, H\'{e}non map expresses the
principal properties of large class of differential systems.

Let us complexify the map~(\ref{hen}) in a such way that
$z,w,\lambda\in\mathbb{C}$, $d\in\mathbb{R}$. According to the
work~\cite{Ipnd,I1}, complexified H\'{e}non map can be
reduced to the two symmetrically coupled real H\'{e}non maps and
can be realized at the physical experiment~\cite{IP}.

Let us remark that with $|d|<1$ the H\'{e}non map is dissipative
system, with $|d| \rightarrow 1$ -- it is area-preserving map, and
with $d \rightarrow 0$ it corresponds to complex 1D quadratic
map~(\ref{log}). 

In previous work~\cite{Ihen} we have investigated intrinsic 
to 1D complex maps special scenario of transition to chaos 
through the period multiplication (for example period-tripling) 
bifurcation cascades and found that accumulation points of these 
bifurcatuins survives for the H\'{e}non map with $1>|d|\neq0$.
In present work we aim to be convinced of existence of one more 
interesting critical phenomena -- so-called Siegel disk for the 
H\'{e}non map.
This special type of critical dynamics of the one-dimensional 
complex analytical maps corresponds to a stability loss of 
the fixed point in case of an irrational winding number, 
that is an irrational phase of a complex multiplier with 
unit modulo $\mu=e^{2\pi i \varphi}$ (for example equal to 
"golden mean" $\varphi =(\sqrt{5}-1) /2$). Siegel disk -- 
is a domain inside Julia set at the phase plane, filled 
with the invariant quasiperiodic trajectories rotating 
around of a neutral fixed point. 

Implementation of domain of rotation about the fixed point 
is connected with existence of local variable change $z=h 
(\omega)$  -- conjugacy function, which conjugate $f$ with 
irrational rotation $\omega\rightarrow\omega\mu$ near to 
the fixed point so, that
\begin{equation}\label{sch}
f(h(\omega))=h(\mu\omega).
\end{equation}
This equation, known as Schr\"{o}der equation, means, that 
each iteration of $f$ is equivalent to rotation of new 
variable $\omega$ on an angle $2\pi\varphi$. Conjugacy 
function (smooth differentiable) exists not at every value 
of complex variable. The domain of its existence corresponds 
to Siegel disk. The boundary of this disk is described by the 
special fractal quasiperiodic trajectory started from an 
extremum of map (images and pre-images of an extremum are 
distributed on this boundary dense everywhere). Really, let 
us take derivative of the Schr\"{o}der equation with regard 
for $h (\omega) =z$
\begin{equation}\label{dsch}
f'(z)h'(\omega)=\mu h'(\mu\omega).
\end{equation}
Let's consider a critical value of the variable $f ' (z_c) =0$. 
From last expression it is evident, that for this value (and 
also for its images and pre-images) conjugacy function should 
have a singularity -- its derivatives should turn to zero or 
infinity.. Besides it is necessary to note, that the extremum 
is the point of a Siegel disk mostly distant from the fixed 
point. 
   
At Fig 1 the phase plane for map~(\ref{log}) with a value of parameter 
\begin{equation}\label{ls}
\lambda^{Siegel}=0.3905409+0.5867879 i,
\end{equation}
relevant to existence of an irrational neutral fixed point 
with a "golden mean" winding number is represented. Several 
invariant curves and boundary of a Siegel disk are shown. 
In a neighbourhood of the fixed point invariant curves are 
close to circles. Invariant curves more distant from the 
fixed point are more distorted. The boundary of a disk, 
defined by a trajectory, started from an extremum is a 
fractal curve.

Renormalization group analysis of dynamics at the Siegel 
disk boundary has been developed by M.Widom~\cite{Widom} 
(see also paper of N.S.~Manton and M.~Nauenberg~\cite{Manton}). 
According to this works, two types of scaling are possible 
for quasiperiodic trajectories at Siegel disk: 

1) So-called simple scaling
\begin{equation}\label{ss}
\frac{f^{F_{N+1}}(z)-z}{f^{F_N}(z)-z}=-\varphi,
\end{equation}
which is valid for any point $z$ inside a disk (Here 
$F_N$ are Fibonacci numbers, and $\varphi$ -- "golden 
mean" winding number); 

2) Boundary of Siegel disk nontrivial scaling law
\begin{equation}\label{ns}
\frac{f^{F_{N+1}}(z_c)-z_c}{f^{F_N}(z_c)-z_c}=\Biggl\{ 
\begin{array}{c}
\alpha,\quad N - \rm even, \\ 
\alpha^*,\quad N - \rm odd, \end{array} 
\end{equation}
with universal scaling factor
\begin{equation}\label{al}
\alpha=-0.22026597-0.70848172i. 
\end{equation}
Let's consider now complex H\'{e}non map. 
The fixed points of map~(\ref{hen}) are
\begin{equation}\label{fpl}
z_{1,2}=w_{1,2}=\frac{-(1+d)\pm\sqrt{(1+d)^2+4\lambda}}{2}. 
\end{equation}

There are two complex multipliers $\mu_{1,2}$ 
for map~(\ref{hen}), which can be found as eigenvalues of a 
Jacobi matrix
\begin{equation}\label{jacobi1}
\mathbf{J}(z,w)={\left( {{\begin{array}{*{20}c}
 {{{\partial f(z,w)}/{\partial z}}} \hfill &
{{{\partial g(z,w)}/{\partial z}}} \hfill \\
 {{{\partial f(z,w)}/{\partial w}}} \hfill &
{{{\partial g(z,w)}/{\partial w}}} \hfill \\
\end{array}} } \right)}=
{\left( {{\begin{array}{*{20}c}
 -2z & -d \\ 1 & 0 \\
\end{array}} } \right)}.
\end{equation}
Thus, multipliers satisfy to the equations 
\begin{equation}\label{mm}
\mu_1\mu_2=Det \mathbf{J}=d,\qquad \mu_1+\mu_2=Tr \mathbf{J}=-2z. 
\end{equation}
One can found the value of parameter $\lambda$, at which 
one of the fixed points~(\ref{fpl}) is neutral
\begin{equation}\label{ln}
\lambda=\frac{1}{4}((e^{2\pi i \varphi}+d e^{-2\pi i \varphi})^2-
2(1+d)(e^{2\pi i \varphi}+d e^{-2\pi i \varphi})). 
\end{equation}
Whereas $\mu_1\mu_2=d$ ($|d|<1$), the fixed points with 
neutral multiplier should have the second multiplier with 
modulo less than unity, i.e. the fixed points should be 
attractive in directions, orthogonal to surface in 4D 
phase space, at which Siegel disk can occur.  

With $d = -0.3$ and $\varphi=(\sqrt{5}-1)/2$ value of 
parameter corresponded to a Siegel disk is
\begin{equation}\label{lll}
\lambda=0.05447888511456006+0.5339769956857415i. 
\end{equation}
The fixed point is placed in a phase space at 
\begin{equation}\label{zw0}
z_0=w_0= 0.25807910732741196 + 0.4390686912699904 i. 
\end{equation}

Nontrivial problem is the determination of the scaling 
center of Siegel disk of H\'{e}non map, which corresponds 
to an extremum in the case of one-dimensional map. For 
this purpose the special original numerical method based 
on universal scaling properties of a Siegel disk~(\ref{ss}) 
and~(\ref{ns}), has been developed. 

The main content of a method is following. One should start 
from any initial point in a multi-dimensional phase space 
close enough to the fixed point~(\ref{zw0}). After several 
time iterations of map~(\ref{hen}) trajectory of this 
point attracts due to a dissipation to one of smooth 
quasiperiodic invariant curves of Siegel disk. Then one 
should determine the point $(z^{(1)},w^{(1)})$ at this invariant 
curve, which is farthest from rotation center~$( z_0,w_0)$. 
One should prolong iterations, until images (of the 
Fibonacci numbers orders) of this point obey with sufficient 
accuracy to a law of simple scaling.
\begin{equation}\label{ss2}
\frac{f^{F_{N+2}}(z^{(1)},w^{(1)})-z^{(1)}}{f^{F_N}(z^{(1)},w^{(1)})-z^{(1)}}=\varphi^2,\qquad
\frac{g^{F_{N+2}}(z^{(1)},w^{(1)})-w^{(1)}}{g^{F_N}(z^{(1)},w^{(1)})-w^{(1)}}=\varphi^2.
\end{equation}
According to the nontrivial scaling law of a Siegel disk boundary 
one can make small step in the direction of scaling center and 
find its approximate location as
\begin{equation}\label{ns2}
\begin{array}{c}
z^{(2)}=\frac{1}{\alpha^2-1}\left(\alpha^2 f^{F_N}(z^{(1)},w^{(1)})-f^{F_{N+2}}(z^{(1)},w^{(1)})\right), \\
w^{(2)}=\frac{1}{\alpha^2-1}\left(\alpha^2 g^{F_N}(z^{(1)},w^{(1)})-g^{F_{N+2}}(z^{(1)},w^{(1)})\right).
\end{array}
\end{equation}
The point $(z^{(2)},w^{(2)})$ belongs to a new invariant 
curve, which is farther from the fixed point and closer 
to the boundary of disk. Then one should repeat procedure: 
1) find farthest from rotation center point of the invariant curve; 
2) determine the scale at which this curve is smouth (invariant 
curves more closer to the disk boundary are more distorted), i.e. 
determine order $F_N$, for which simple scaling implements; 3) 
calculate new approximation of the boundary scaling center using 
position of $F_N$ and $F_{N+2}$ -th iterations of farthest point.

Repeating procedure more and more times the position of scaling center 
can be calculated more and more precisely 
($(z^{(n)},w^{(n)})\rightarrow (z_c,w_c)$ with $n\rightarrow \infty$).
As a result the following position of scaling center have been found 
\begin{equation}\label{zw}
\begin{array}{c}
z_c=0.45756301999 + 0.30965877i, \\
w_c= 0.22028610776 + 0.64312280i.
\end{array}
\end{equation}

At Fig.~2 the projection of Siegel disk to three-dimensional 
space is represented. Siegel disk looks as smouth surface 
with fractal boundary. Several invariant curves (concentric 
lines) on a disk are represented. Radial 
lines at figure correspond to images of a curve, 
which is a locus of invariant curves points mostly distant from the 
fixed point. Scaling center of Siegel disk is designated.

\vspace{7mm}
\it
The work is performed under partial supoport of RFBR (Grant No~06-02-16619), 
O.B.I. acknowledges support from INTAS (Grant 05-109-5262).
\rm

\newpage

\begin{figure}
\centerline{\epsfig{file=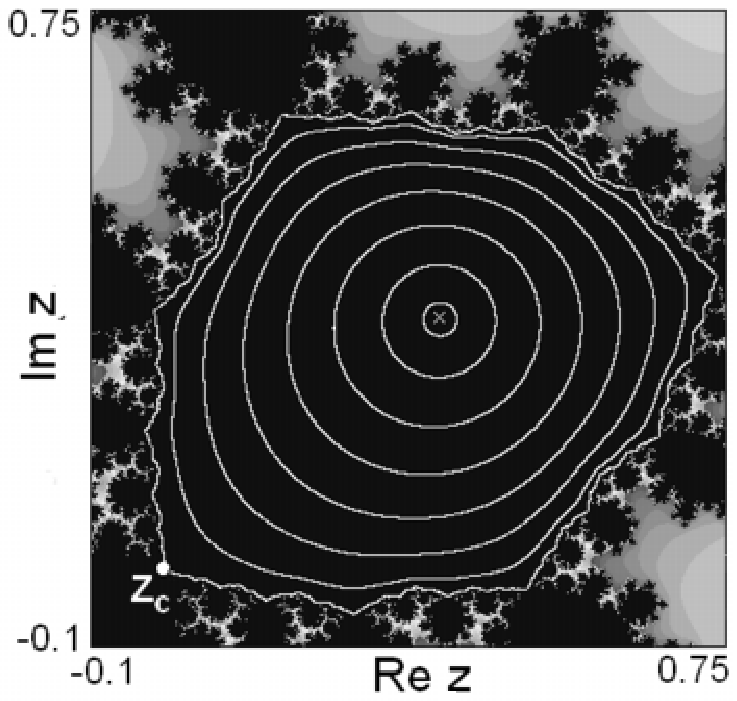, width=0.4\textwidth}}

\caption{Phase plane of complex map~(\ref{log}). 
By gray colors the areas of starting values of 
variable, which iterations escape to infinity by 
different time are marked. Black color corresponds 
to restricted in a phase space dynamics. Curves 
represent quasiperiodic trajectories around of a 
neutral fixed point (designated by a obelisk). The fractal 
trajectory started from an extremum of map (designated 
by a circle) represents boundary of a Siegel disk.}
\end{figure}
\begin{figure}
\centerline{\epsfig{file=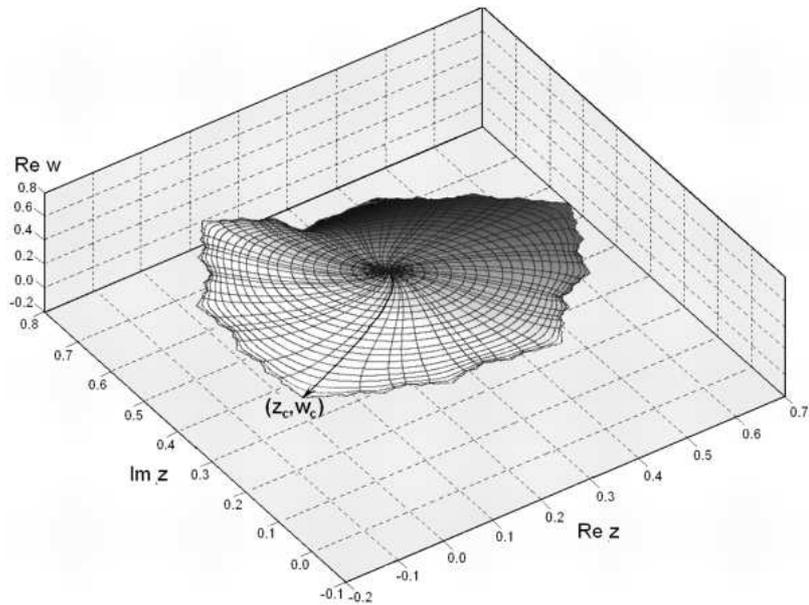,width=0.6\textwidth}}

\caption{A projection of a Siegel disk for the 
complexified H\'{e}non map~(\ref{hen}) to a 
three-dimensional phase space. Concentric lines on 
a disk represent quasiperiodic trajectories. Radial 
lines correspond to images of a curve (bold line with arrow), 
which is a locus 
of invariant curves points mostly distant from the 
fixed point. Scaling center of Siegel disk (the point 
being analog of an extremum) is designated as $(z_c,w_c)$.}
\end{figure}


\newpage

\begin {thebibliography}{9}
\bibitem{Peitgen} H.-O. Peitgen, P.H. Richter. The beauty of fractals.
Images of complex dynamical systems. Springer-Verlag. 1986.

\bibitem{Beck} C.~Beck. Physical meaning for Mandelbrot and Julia set. //
Physica D125, 1999, P. 171-182.

\bibitem{Ipnd} O.B.~Isaeva. // Izv. VUZov PND (Appl. Nonlin. Dyn.), 
V. 9, No. 6, 2001, P. 129-146 (in Russia).

\bibitem{IP} O.B.~Isaeva, S.P.~Kuznetsov, V.I. Ponomarenko.
Mandelbrot set in coupled logistic maps and in an electronic
experiment. // Phys. Rev. E. V. 64, 2001, 055201(R).

\bibitem{I1} O.B.~Isaeva, S.P.~Kuznetsov. On possibility of 
realization of the phenomena of complex analytic dynamics in 
physical systems. Novel mechanism of the synchronization loss 
in coupled period-doubling systems.// http://www.arXiv.org/abs/nlin.CD/0509012.

\bibitem{I2} O.B.~Isaeva, S.P.~Kuznetsov. On possibility of 
realization of the Mandelbrot set in coupled continuous systems. 
// http://www.arXiv.org/abs/nlin.CD/0509013.

\bibitem{Kuznetsov} S.P.~Kuznetsov. Dynamical chaos. M.: Nauka. 2001 (in Russia).

\bibitem{Ihen} O.B.~Isaeva, S.P.~Kuznetsov. Period tripling accumulation 
point for complexified H\'{e}non map. // http://www.arXiv.org/abs/nlin.CD/0509015.

\bibitem{Widom} M.~Widom. Renormalization group analysis of quasi-periodicity in
analytic maps. // Commun. Math. Phys., V. 92, 1983, P. 121-136.

\bibitem{Manton} N.S.~Manton, M.~Nauenberg. Universal scaling bahavior for
iterated maps in the complex plane. // Comm. Math. Phys., V. 89,
1983, P. 557.


\end{thebibliography}
\end{document}